\documentclass[prl,aps,twocolumn,floatfix,citeautoscript,showpacs]{revtex4-2}
\usepackage{graphicx,rotating,subfigure,amsmath,amsfonts,amssymb,delarray,color}
\usepackage{hyperref}
\usepackage{xcolor}
\usepackage{soul}
\usepackage{dsfont}
\usepackage[T1]{fontenc}
\usepackage{physics}
\usepackage{multirow}
\usepackage{bbm}

\def\12{\frac{1}{2}}

\begin{document}
\title{Symmetry-Resolved Entanglement of $C_2$-symmetric Topological Insulators}
\author{Kyle Monkman}
\author{Jesko Sirker}
\affiliation{Department of Physics and Astronomy and Manitoba
Quantum Institute, University of Manitoba, Winnipeg, Canada R3T 2N2}
\date{\today}
\begin{abstract}
For a many-body system of arbitrary dimension, we consider fermionic ground states of non-interacting Hamiltonians invariant under a $C_2$ cyclic group. The absolute difference $\Delta$ between the number of occupied symmetric and anti-symmetric single-particle states is an adiabatic invariant. We prove lower bounds on the configurational and the number entropy based on this invariant. In band insulators, the topological invariant $\Delta$ and the entropy bounds can be directly determined from high symmetry points in the Brillouin zone. 
\end{abstract}
\maketitle
%\section{Introduction}
\paragraph*{Introduction.---$\!\!\!\!$} Symmetry-resolved entanglement is a term for entanglement measures which take into account superselection rules. The superselection rules dictate properties of the system such as the amount of useful entanglement which can be extracted and transferred to a quantum register \cite{Wiseman}. An important example is the superselection of particle number because entanglement is essentially only extractable after a measurement of the local particle number. Bounds on the symmetry-resolved quantities are important to determine a minimum or maximum usefulness of a quantum state. In addition, symmetry-resolved entanglement has been used to study many-body physics \cite{KieferSirker1, KieferSirker2, Greiner, Parez, Bonsignori, Sela}, quantum field theories \cite{field1, field2, field3, field4, field5}, and topological systems \cite{MonkmanSirkerEdge, topology1, topology2}.

A topological invariant is an ideal candidate for finding lower bounds on entanglement measures since topology is robust to symmetry-conserving deformations \cite{crystalline, inversion, tenfold1, three, tenfold2}. The single-particle entanglement spectrum of a symmetry-protected topological phase has been studied for point-group symmetries in Refs.~\cite{Bernevig2011, Bernevig2013, Bernevig2014} and then later for chiral symmetry in Ref.~\cite{MonkmanSirkerSpectrum}. The properties of the spectrum imply a non-zero lower bound on the von-Neumann entanglement entropy in such a phase. Furthermore, numerical evidence of topologically protected entanglement bounds has been found for a wide variety of systems \cite{Poyhonen}. 

Bounds on the symmetry-resolved entanglement can enable a more complete picture of topological protection. In this article, we prove lower symmetry-resolved entanglement bounds for ground states of fermionic systems invariant under a $C_2$ cyclic group which includes, for example, the point groups of inversion, $180^\circ$ rotation, and a mirror plane reflection. We obtain lower bounds for two types of symmetry-resolved entanglement: the configurational entropy and the number entropy. Using the absolute difference $\Delta$ between occupied symmetric and anti-symmetric single-particle states as a topological invariant, we find that the bounds can be understood as the configurational and the number entropy of $\Delta$ independent dimers.

Our article is organized as follows: We first introduce the concept of $C_2$ symmetry, define an adiabatic invariant, and prove a preliminary result. Next, we prove lower symmetry-resolved entanglement bounds on Gaussian states for general Hamiltonians with $C_2$ symmetry. As an example, we consider the case of a one-dimensional, inversion-symmetric band insulator and show that in this case the topological invariant can be obtained from the occupation of the bands at the inversion-symmetric momenta. Finally, we will conclude and summarize our results.

%\section{Model and preliminary results}
%\label{section2}
\paragraph*{Model and preliminary results.---$\!\!\!\!$}In this article, we only deal with fermions and we use the notation of second quantization. Fermionic creation operators $c_i^\dag$ with $\{c_i,c_j^\dag\}=\delta_{ij}$ create the state $|c_i \rangle=c_i^\dag |0 \rangle$ where $|0\rangle$ is the vaccuum state. Multi-particle states are denoted, for example, as $|c_i \rangle |c_j \rangle = c_i^\dag c_j^\dag |0\rangle$.

We will consider a system in any dimension consisting of a subsystem $L$ and a subsystem $R$ with an equal number of single particle states, $M_L=M_R$, in each subsystem. We denote single-particle states with one particle in subsystem $L$ by $|\ell_i\rangle \equiv |\ell_i\rangle\otimes |0_r\rangle$ and the ones with one particle in subsystem $R$ by $|r_j\rangle \equiv |0_\ell\rangle\otimes |r_j\rangle$. This is to indicate that these single particle states are localized in either subsystem. We choose these states such that they form an orthonormal basis for their subsystem. We denote an orthonormal basis of the entire system by $|c_i \rangle$. This basis has dimension $2M_L=2M_R$. 

\paragraph*{Definition 1:} Let $\hat I$ be a conserved operator which is unitary and Hermitian and which thus fulfills $\hat{I}^2=\mathbbm{1}$. This means that $\hat I$ is a generator of the cyclic group $C_2$ which is the unique group of group order $2$. Examples are the point groups of inversion, mirror symmetry, and $180^\circ$ degree rotation but our approach is general and also applies to any inner $C_2$ symmetry. We can define the action of the operator $\hat I$ by 
\begin{equation}
\label{Definition2}
\hat{I} |\ell_i \rangle = \sum_j I_{i,j} | r_j \rangle \ , \ \hat{I} |r_i \rangle = \sum_j I_{i,j} | \ell_j \rangle
\end{equation}
where the matrix $I_{i,j}$ is unitary and Hermitian. A state $| c_i \rangle$ is symmetric with respect to $\hat{I}$ if $\hat{I} | c_i \rangle = | c_i \rangle $. Similarly, a state $| c_i \rangle$ is anti-symmetric with respect to $\hat{I}$ if $\hat{I} | c_i \rangle = -| c_i \rangle $. We note that because $\hat{I}$ has the property $\hat{I}^2=\mathbbm{1}$, it can be written as $\hat{I}=2\hat{P}-\mathbbm{1}$. Here $\hat{P}$ is a projection operator which projects onto the symmetric part of the state. Also, since $I_{i,j}$ is unitary, the states $\{\hat{I} | \ell_i \rangle\}$ constitute a basis for the subsystem $R$. If $| c_i \rangle$ is a symmetric state, then it can be written as $|c_i\rangle = (\frac{1+\hat{I}}{\sqrt{2}})|c_i^\ell\rangle$, where $|c_i^\ell\rangle$ is a state in the $L$ subspace $\lbrace |\ell_i \rangle \rbrace$. Similarly, if $| c_i \rangle$ is anti-symmetric then it can be written as $|c_i\rangle = (\frac{1-\hat{I}}{\sqrt{2}})|c_i^\ell\rangle$. 

We will consider fermionic many-particle ground states $|\Psi \rangle$ of non-interacting, $C_2$-symmetric Hamiltonians $\hat{H}=\hat{I} \hat{H} \hat{I}$, with $S$ symmetric, and $A$ anti-symmetric filled single-particle states. The total number of filled single particle states is thus $M=A+S$. We are interested in systems where there is an energy gap between the occupied and unoccupied states. If we consider adiabatic deformations $\hat{H} \rightarrow \hat{H}+\hat{V}$ which respect the $C_2$ symmetry, i.e. $\hat{I} \hat{V} \hat{I} =\hat{V}$, then symmetric and antisymmetric states never couple. Using, in addition, that the energy gap is maintained in such an adiabatic transformation it follows that the quantities $A$ and $S$ are adiabatic invariants. Since $A$ and $S$ often scale with system size, we use the absolute difference $\Delta=|S-A|$ as the relevant adiabatic invariant.

In these ground states, we can always find single-particle states $|a_i^\ell \rangle$ and $|s_i^\ell \rangle$ in the subspace $L$ with $\langle a_i^\ell |a_j^\ell \rangle =\delta_{ij}$ and $\langle s_i^\ell |s_j^\ell \rangle =\delta_{ij}$ such that
\begin{equation}
\label{Psi}
|\Psi \rangle = \prod_{i=1}^{S} (\frac{1+\hat{I}}{\sqrt{2}})|s_i^\ell \rangle \prod_{j=1}^{A} (\frac{1-\hat{I}}{\sqrt{2}})|a_j^\ell \rangle \, .
\end{equation}
Importantly, $\langle a_i^\ell |s_j^\ell \rangle\neq \delta_{ij}$ in general, see below. It is important to note that if we replace $|s_j^\ell \rangle$  with a unitary transformed basis $|\tilde s_j^\ell \rangle$ then $|\Psi \rangle$ remains invariant up to a phase. Let $\mathbb{L}$ be the vector space of the combined span of the $|s_i^\ell \rangle$ and $|a_i^\ell \rangle$ vectors. Furthermore, we define $\mathbb{S}=\mbox{span}\{|s_i^\ell \rangle\}$ and $\mathbb{A}=\mbox{span}\{|a_i^\ell \rangle\}$. Then since $\mathbb{S}$ and $\mathbb{A}$ are subspaces of $\mathbb{L}$, we can define complement vector spaces $\mathbb{S}^\bot=\mathbb{L} \ / \ \mathbb{S}$  and $\mathbb{A}^\bot=\mathbb{L} \ / \ \mathbb{A}$. 

If $S>A$, then
$S = \text{dim}(\mathbb{S}\cap\mathbb{A})+ \text{dim}(\mathbb{S}\cap\mathbb{A}^\bot) = A+\Delta$.
Using $\text{dim}(\mathbb{S}\cap\mathbb{A}) \leq A$, we also have $\Delta \leq \text{dim}(\mathbb{S}\cap\mathbb{A}^\bot)$. This is important because it means that we can always find at least $\Delta$ states $|\tilde s_i^\ell \rangle$ such that $\langle \tilde s_i^\ell | a_j^\ell \rangle=0$ for $i\in\{1,\dots,\Delta\}$ and $j\in\{1,\dots,A\}$. Forming an orthonormal set $\{|\tilde s^\ell_i\rangle\}$ using these states we then define
\begin{eqnarray}
\label{D}
|D\rangle &=& \prod_{i=1}^{\Delta} (\frac{1+\hat{I}}{\sqrt{2}})|\tilde s_i^\ell \rangle \nonumber \\
|\phi \rangle &=& \prod_{i=\Delta+1}^{S} (\frac{1+\hat{I}}{\sqrt{2}})|\tilde s_i^\ell \rangle \prod_{i=1}^{A} (\frac{1+\hat{I}}{\sqrt{2}})|a_i^\ell \rangle 
\end{eqnarray}
so that $|\Psi \rangle = |D\rangle |\phi \rangle$. Note that $|D \rangle$ is essentially a product of $\Delta$ independent dimers.

If, on the other hand $A>S$, we can still define states $|D \rangle$ and $|\phi \rangle$ by swapping $|\tilde s_i^\ell \rangle$ terms with $|\tilde a_i^\ell \rangle$ terms and vice versa. In either case, $|D\rangle$ still represents $\Delta$ independent dimers and $|\Psi \rangle = |D\rangle |\phi \rangle$.

\paragraph*{Entanglement of $C_2$-symmetric Gaussian states.---$\!\!\!\!$} For any normalized state $|\psi \rangle$, let $S[ |\psi \rangle ]$ be the von-Neumann entanglement entropy of the subsystem $L$. Let $\hat{P}_n$ be the projection operator onto states with $n$ particles in this subsystem. The probability of having $n$ particles in the subsystem is given by $P_n^\psi = \langle \psi | \hat{P}_n | \psi \rangle$. The normalized projection is $|\psi_n \rangle = \frac{1}{\sqrt{P_n^\Psi}} \hat{P}_n |\psi \rangle$. The configurational entropy $S_C$ and number entropy $S_N$ are then defined as
\begin{eqnarray}
S_C[|\psi \rangle] &=& \sum_n P_n^\psi S[|\psi_n \rangle] \nonumber \\
S_N[|\psi \rangle] &=& -\sum_n P_n^\psi \ln P_n^\psi 
\end{eqnarray}
where the summation index $n$ is over all possible particle numbers. For a general state $|\psi \rangle $, we will utilize the well known equality \cite{Wiseman,Greiner}
\begin{equation}
\label{equal}
S[|\psi \rangle ] = S_C[|\psi \rangle ]+S_N[|\psi \rangle ].
\end{equation}
We will now consider the states $|\phi \rangle$, $|D \rangle$ and $|\Psi \rangle = |D\rangle | \phi \rangle$, defined in Eq.~\eqref{D}. The entropies of the state $|D\rangle$ are calculable straightforwardly, and we will show that they are a lower bound for the entropies of the state $|\Psi\rangle$. Let $C^n_r \equiv \begin{pmatrix} n \\ r \end{pmatrix}$ be a binomial coefficient. Then $P_r^D=C_r^\Delta/2^\Delta$ and the entanglement measures of the state $|D\rangle$ are
\begin{eqnarray}
\label{results}
S_N[|D \rangle] &=& - \sum_{r=0}^{\Delta} \frac{C^\Delta_r}{2^\Delta} \ln(\frac{C^\Delta_r}{2^\Delta}) \nonumber \\
S_C[|D\rangle] &=& \sum_{r=0}^\Delta \frac{C^\Delta_r}{2^\Delta} \ln(C^\Delta_r) \\
S[|D\rangle] &=& \Delta \ \ln 2. \nonumber
\end{eqnarray}

\paragraph*{Result 1:} Suppose a state $|\Psi \rangle=|D\rangle|\phi\rangle$ is of the form \eqref{D}. Then 
\begin{equation}
\label{Result2}
S_C[|\Psi \rangle ] \geq S_C[|D\rangle].
\end{equation}

\paragraph*{Proof of Result 1:} If we expand the products in $|D\rangle$, we obtain a sum of $2^\Delta$ states. In order to have $r$ particles in subsystem $L$ we have to pick $r$ states $|\tilde s_i^\ell\rangle$ and $\Delta-r$ states $\hat{I}|\tilde s_i^\ell\rangle$. There are thus $C^\Delta_r$ of such states. We will label these states as $|D_r(j) \rangle$ for $1 \leq j \leq C^\Delta_r$. Then we can write $|D\rangle$ as 
\begin{equation}
|D \rangle = \frac{1}{\sqrt{2^{\Delta }}} \sum_{r=0}^\Delta \sum_{j=1}^{C^\Delta_r} | D_r (j) \rangle
\end{equation}
with $\langle D_r(j)|D_{r'}(j')\rangle =\delta_{rr'}\delta_{jj'}$.
Defining $n'=\text{min}(n,\Delta)$ we can now write
\begin{eqnarray}
|\Psi_n \rangle &=& \frac{1}{\sqrt{P_n^\Psi}}\hat P_n |D\rangle|\phi\rangle = \frac{1}{\sqrt{P_n^\Psi}}\sum_{r=0}^{n'}\hat P_r|D\rangle \hat P_{n-r}|\phi\rangle \nonumber \\
&=& \frac{1}{\sqrt{P_n^\Psi 2^\Delta}} \ \  \sum_{r=0}^{n'} \sqrt{P_{n-r}^\phi} \sum_{j=1}^{C^\Delta_r} |D_r (j) \rangle |\phi_{n-r} \rangle
\end{eqnarray}
with $|\phi_{n-r}\rangle=\frac{1}{\sqrt{P^\phi_{n-r}}}\hat P_{n-r}|\phi\rangle$.

We will define reduced density matrices $\rho_{n}^\Psi = \text{tr}_R[|\Psi_n \rangle \langle \Psi_n |]$, $\rho_{n}^\phi = \text{tr}_R[|\phi_n \rangle \langle \phi_n |]$, and $\rho_{r,j}^D = \text{tr}_R[| D_r (j) \rangle \langle D_r (j) |]$, where the $\text{tr}_R[ \ . \ ] $ operation traces out subsystem $R$. Due to the dimer property of $|D\rangle$, $\rho_{r,j}^D$ is a pure state density matrix and a basis exists where $\rho_{r,j}^D$ has a single one along the diagonal and zeros in all other entries. Furthermore, each of the $\rho_{r,j}^D$ have a different diagonal element with a one. For example, in this basis, we have $(\rho_{r,j}^D)^2 = \rho_{r,j}^D$ and $\rho_{r,j}^D \rho_{r',j'}^D = \delta_{r,r'} \delta_{j,j'} \rho_{r,j}^D$. Then, we find
\begin{equation}
\rho_{n}^{\Psi}
= \sum_{r=0}^{n'} \sum_{j=1}^{C^\Delta_r} \frac{P_{n-r}^\phi}{P_n^\Psi 2^\Delta} \  \rho_{r,j}^D \otimes \rho_{n-r}^\phi . 
\end{equation}
Note that $\rho_{r,j}^D \otimes \rho_{n-r}^\phi$ is a block matrix with one block given by $\rho_{n-r}^\phi$ while all other blocks are zero. Only the position of the $\rho_{n-r}^\phi$-block depends on $j$. We can therefore write the von-Neumann entanglement entropy as
\begin{eqnarray}
\label{S1}
&& S[|\Psi_n \rangle] = -\text{tr}[ \rho_n^\Psi \ln \rho_n^\Psi ] \nonumber \\
&&=\sum_{r=0}^{n'} \frac{P_{n-r}^\phi \ C^\Delta_r}{P_n^\Psi \ 2^\Delta} \left\{ \ \ln\left(\frac{P_n^\Psi \ 2^\Delta}{P_{n-r}^\phi}\right) -\text{tr}[\rho_{n-r}^\phi \ln(\rho_{n-r}^\phi)]  \ \right\} \nonumber \\
&&\geq \sum_{r=0}^{n'} \frac{P_{n-r}^\phi \ C^\Delta_r}{P_n^\Psi \ 2^\Delta} \ln\left(\frac{P_n^\Psi \ 2^\Delta}{P_{n-r}^\phi}\right). 
\end{eqnarray}
Using the inequality
\begin{equation}
\label{S2}
\frac{P_n^\Psi}{P_{n-r}^\phi} = \frac{\sum_{j=0}^{n'} P_j^D P_{n-j}^\phi}{P_{n-r}^\phi} \geq P_r^D =\frac{C^\Delta_r}{2^\Delta}
\end{equation}
we find
\begin{equation}
\label{S3}
S[|\Psi_n \rangle] \geq 
\sum_{r=0}^{n'} \frac{P_{n-r}^\phi \ C^\Delta_r}{P_n^\Psi \ 2^\Delta} \ln(C^\Delta_r).
\end{equation}
Then
\begin{eqnarray}
S_C [ |\Psi \rangle ] &=& \sum_{n=0}^{M} P_n^\Psi S[ |\Psi_n \rangle] \\ 
&\geq& 
\sum_{r=0}^{\Delta} \sum_{n=r}^{M} \frac{P_{n-r}^\phi \ C^\Delta_r}{2^\Delta} \ln(C^\Delta_r) = \sum_{r=0}^\Delta \frac{C^\Delta_r}{2^\Delta} \ln(C^\Delta_r). \nonumber
\end{eqnarray}
This proves the inequality \eqref{Result2}.
\hfill $\blacksquare$ \\ 

\paragraph*{Result 2:} Suppose a state $|\Psi \rangle=|D\rangle|\phi\rangle$ is of the form \eqref{D}. Then 
\begin{equation}
\label{Result3}
S_N[|\Psi \rangle ] \geq S_N[|D\rangle].
\end{equation}

\paragraph*{Proof of Result 2:} 
We need to show that 
\begin{equation}
\label{SN}
    S_N[|\Psi\rangle] = -\sum_{n=0}^M P_n^\Psi\ln P_n^\Psi\geq -\sum_{r=0}^\Delta P_r^D\ln P_r^D
\end{equation}
with $M=S+A$, $P_r^D=C_r^\Delta/2^\Delta$ and $P_n^\Psi=\sum_{r=0}^{n'} P_r^D P_{n-r}^\phi$. To do so, we will use Karamata's inequality \cite{Karamata1, Karamata2,karamata3}. First, we note that $g(P)=-P\ln P$ is a concave function. We also note that in the following $P_i^D$, $P_i^\phi$ and $P_i^\Psi$ are considered to be zero if the index $i$ is greater than or less than the number of allowed particles in $|D\rangle$, $|\phi \rangle$ and $|\Psi \rangle$ respectively. Let $\sigma(i)$ and $\alpha(i)$ be permutations of $(0, \ 1, \  \dots \ M)$ such that
$x_0=P_{\sigma(0)}^D\geq x_1=P_{\sigma(1)}^D\geq\dots\geq x_M=P_{\sigma(M)}^D$ and $y_0=P_{\alpha(0)}^\Psi\geq y_1=P_{\alpha(1)}^\Psi\geq\dots\geq y_M=P_{\alpha(M)}^\Psi$.

Now, if we can show that $X$ majorizes $Y$, then \eqref{SN} is true by Karamata's inequality.  We do know that $\sum_{i=0}^M x_i = \sum_{i=0}^M y_i = 1$. We need to show that $\sum_{i=0}^s y_i \leq \sum_{i=0}^s x_i$ for all $s \in \lbrace 0, 1, \dots M-1 \rbrace$. So
\begin{eqnarray}
&&\sum_{i=0}^s y_i = \sum_{i=0}^s P_{\alpha(i)}^\Psi 
=\sum_{i=0}^s \sum_{j=0}^{\alpha(i)} P_j^\phi P_{\alpha(i)-j}^D \\
&=&\sum_{j=0}^M P_j^\phi (\sum_{i=0}^s P_{\alpha(i)-j}^D)  \leq 
\sum_{j=0}^M P_j^\phi (\sum_{i=0}^s x_i)  = \sum_{i=0}^s x_i. \nonumber
\end{eqnarray}
Thus $X$ majorizes $Y$ and \eqref{SN} is true by Karamata's inequality.
\hfill $\blacksquare$ \\ 

\paragraph*{Result 3:} Suppose a state $|\Psi \rangle=|D\rangle|\phi\rangle$ is of the form \eqref{D}. Then 
\begin{equation}
S[|\Psi \rangle ] \geq S[|D \rangle ].
\end{equation}

\paragraph*{Proof of Result 3:} Using \eqref{equal}, this follows directly from Results 2 and 3. \hfill $\blacksquare$  

%\section{Inversion-symmetric band insulators}
%\label{section4}
\paragraph*{Inversion-symmetric band insulators.---$\!\!\!\!$}As an example, we want to apply the results derived previously, which are valid for any $C_2$ symmetry in any dimension, to the specific case of a one-dimensional insulator with inversion symmetry. We will discuss, in particular, how the invariant $\Delta$ can be determined directly from the high symmetry momenta $k_{\rm inv}=0,\pi$. 

We consider a one-dimensional, periodic system with an even number $L$ of unit cells and $N$ elements per unit cell. Let $c_j^n$ be an annihilation operator on lattice site $j$ of type $n$. Let $c_k^n = \frac{1}{\sqrt{L}} \sum_j e^{-ikj} c_j^n$ where the $k$ values are integers multiplied by $\frac{2 \pi}{L}$. We set ${\psi_j}^\dag = \begin{pmatrix} {c_j^1}^\dag \ , \ {c_j^2}^\dag \ , \ \dots \ , \ {c_j^N}^\dag \end{pmatrix}$ and ${\psi_k}^\dag = \begin{pmatrix} {c_k^1}^\dag \ , \ {c_k^2}^\dag \ , \ \dots \ , \ {c_k^N}^\dag \end{pmatrix}$ and we define the inversion operator $\hat{I}$ as 
\begin{equation}
\hat{I} =  
\sum_{j=0}^{L-1} {\psi_j}^\dag U_I {\psi_{L-1-j}} = 
\sum_k e^{ik} {\psi_k}^\dag U_I {\psi_{-k}}
\end{equation}
where $U_I$ is an $N \times N$ unitary and Hermitian matrix. $\hat{I}$ maps states between the first $\frac{NL}{2}$ lattice sites and the second $\frac{NL}{2}$ lattice sites. It is important to note that the operator $\hat{I}$ defined here is an example of a generator of the $C_2$ cyclic group as defined in \textit{Definition 1}. 

We will consider a non-interacting Hamiltonian 
\begin{equation}
\hat{H}=\sum_k {\psi_k}^\dag H(k) \psi_k
\end{equation} 
where $H(k)$ is an $N \times N$ Hermitian matrix. In particular, we will consider insulators with Hamiltonians $\hat{H}$ fulfilling $\hat{I} \hat{H} \hat{I} = \hat{H}$. This also implies that that $U_I^\dag H(k) U_I = H(-k)$. Since the Hamiltonian is inversion symmetric and has a band gap, the number $S$ of symmetric states, the number $A$ of anti-symmetric states, and the absolute difference $\Delta=|S-A|$ are all adiabatic invariants. As mentioned earlier, $S$ and $A$ scale with system size, and so we focus on the $\Delta$ invariant. The results for the bounds given in \textit{Result 1}, \textit{Result 2} and \textit{Result 3} now apply to the insulating ground state $|\Psi \rangle$. Thus we have topologically protected configurational, number, and von-Neumann entropy for an inversion-symmetric, insulating state.

Lastly, we will discuss how $\Delta$ can be determined from the high symmetry momenta in the Brillouin zone. Let $|u_k^\ell \rangle$ be a Bloch state of momentum $k$ and band index $\ell$. In terms of creation operators we can write $|u_k^\ell \rangle =\psi_k^\dag u_k^\ell |0\rangle$, where $u_k^\ell$ is an $N$ dimensional vector and $|0\rangle$ is the vacuum state. Then an insulating ground state can be written as
\begin{equation}
\label{bands}
|\Psi \rangle = \prod_{k,\ell} |u_k^\ell \rangle.
\end{equation}
The index $\ell$ runs over a set of lowest energy bands. 

Notice that the filled states $\hat{I} |u_k^\ell \rangle$ and $|u_{-k}^\ell \rangle$ are unitarily equivalent. Therefore, up to some phase factor $z$, we have
\begin{eqnarray}
\label{bands2}
|\Psi \rangle &=& z\prod_{\ell} \left( |u_0^\ell \rangle |u_\pi^\ell \rangle \prod_{0<k<\pi} |u_k^\ell \rangle \hat{I} |u_k^\ell \rangle \right) \\ &=& z\prod_{\ell} \left(|u_0^\ell \rangle |u_\pi^\ell \rangle \prod_{0<k<\pi} (\frac{1-\hat{I}}{\sqrt{2}})|u_k^\ell \rangle (\frac{1+\hat{I}}{\sqrt{2}})|u_k^\ell \rangle \right). \nonumber 
\end{eqnarray}

Therefore, the only contribution to $\Delta=|S-A|$ comes from the inversion symmetric momenta $k=0$ and $k=\pi$. Let $n^{+} (k_{\rm inv})$ ($n^{-} (k_{\rm inv})$) be the number of filled states at inversion symmetric momenta $k_{\rm inv}=0, \ \pi$ with an eigenvalue of $\hat{I}$ equal to $+1$ ($-1$). Then, in line with Refs. \cite{Bernevig2011, Bernevig2013, Bernevig2014}, the topological invariant is given by
\begin{equation}
\Delta = \Biggl\lvert \sum_{k_{\rm inv}} \left[ n^+ (k_{\rm inv}) - n^- (k_{\rm inv})\right] \Biggr\rvert.
\end{equation}
%\section{Conclusion}
%\label{section5}
\paragraph*{Conclusions.---$\!\!\!\!$} Our results are concerned with the amount of useful entanglement contained in a many-body system with particle number conservation. As Wiseman and Vaccaro \cite{Wiseman} have argued, independent measurements and manipulations on each subsystem must be allowed to fully use the shared entanglement. This, however, is only consistent with particle number conservation if the state locally has a definite particle number. The useful entanglement is thus the one obtained after projecting onto a local state with fixed particle number which is usually called the configurational entanglement. Such entanglement is typically fragile to local perturbations in a many-body system unless it is (partially) protected by a topological invariant. Here we have shown, in particular, that there is a finite amount of configurational entanglement which is protected in a topological phase of an insulating state with a $C_2$ symmetry. For a concrete model, this protection can often be understood by studying limiting cases. An example is the Su-Schrieffer-Heeger chain, which is inversion symmetric, in the limit of strong dimerization \cite{Ladder}.

While the configurational entanglement is hard to measure experimentally, the number entanglement is based on the probability distribution of particles alone and can, for example, be obtained straightforwardly in experiments on cold-atomic gases in optical lattices with single-site resolution \cite{Greiner}. As we have shown here, the number entropy is also topologically protected and can thus be used---either  experimentally or in numerical calculations---as an order parameter to detect topological phase transitions.    

The single particle entanglement spectrum of inversion symmetric insulators has been studied in \cite{Bernevig2011, Bernevig2013, Bernevig2014}. This spectrum determines a lower bound for the von-Neumann entanglement of the system. Here we have generalized these results to any type of Gaussian state with $C_2$ symmetry and have obtained lower bounds both for the configurational entropy, relevant for quantum information applications, and for the experimentally accessible number entropy. For the future, it would be interesting to generalize these results to other types of symmetry protected topological order which would allow to obtain a more thorough understanding of the useful entanglement contained in such systems.

\acknowledgments
The authors acknowledge support by the Natural Sciences and Engineering Research Council (NSERC, Canada). K.M. acknowledges support by the Vanier Canada Graduate Scholarships Program. J.S. acknowledges by the Deutsche Forschungsgemeinschaft (DFG) via Research Unit FOR 2316.

%\bibliography{references}

\end{document}